\documentclass[journal]{IEEEtran}

\ifCLASSINFOpdf
\else
\fi

\usepackage[cmex10]{amsmath}

\usepackage{graphicx,epic,eepic,epsfig,amsmath,latexsym,amssymb,mathrsfs,verbatim,subfigure,color}

\usepackage{MnSymbol}
\usepackage{cite}
\usepackage{url}

\newtheorem{definition}{Definition}[section]
\newtheorem{proposition}[definition]{Proposition}
\newtheorem{lemma}[definition]{Lemma}

\newtheorem{theorem}[definition]{Theorem}
\newtheorem{corollary}[definition]{Corollary}

\newtheorem{example}[definition]{Example}

\def\squareforqed{\hbox{\rlap{$\sqcap$}$\sqcup$}}
\def\qed{\ifmmode\squareforqed\else{\unskip\nobreak\hfil
\penalty50\hskip1em\null\nobreak\hfil\squareforqed
\parfillskip=0pt\finalhyphendemerits=0\endgraf}\fi}
\def\endenv{\ifmmode\;\else{\unskip\nobreak\hfil
\penalty50\hskip1em\null\nobreak\hfil\;
\parfillskip=0pt\finalhyphendemerits=0\endgraf}\fi}

\def\wgt{\mathop{\rm wgt}}
\long\def\ignore#1{}

\begin{document}

\title{New Constructions of Codes for Asymmetric Channels via Concatenation}

\author{Markus Grassl,~\IEEEmembership{Member,~IEEE,} Peter Shor, Graeme Smith, John Smolin, and Bei Zeng
\thanks{G. Smith and J. Smolin acknowledge support from the DARPA QUEST program
under contract number HR0011-09-C-0047. B. Zeng is supported by NSERC and CIFAR. The Centre for Quantum Technologies is a Research Centre of Excellence funded by the Ministry of Education and the National Research Foundation of Singapore.}
\thanks{M. Grassl is with the Centre for Quantum Technologies,
  National University of Singapore, Singapore 117543, Republic of
  Singapore (e-mail: Markus.Grassl@nus.edu.sg).}%
\thanks{P. W. Shor is with the Department of Mathematics, Massachusetts
  Institute of Technology, Cambridge MA 02139, USA.}%
\thanks{G. Smith and J. Smolin are with the IBM T. J. Watson Research
  Center, Yorktown Heights, NY 10598, USA.}%
\thanks{B. Zeng is with the Department of Mathematics $\&$ Statistics,
University of Guelph, Guelph, ON, N1G 2W1, Canada and with the
Institute for Quantum Computing, University of Waterloo, Waterloo,
Ontario, N2L 3G1, Canada.}
\thanks{Part of this work has been presented \cite{GSSSZ12} at the 2012 IEEE
  International Symposium on Information (ISIT 2012) in Cambridge, MA, USA.}
}

\maketitle

\maketitle

\begin{abstract}
We present new constructions of codes for asymmetric channels for both
binary and nonbinary alphabets, based on methods of generalized code
concatenation.  For the binary asymmetric channel, our methods
construct nonlinear single-error-correcting codes from ternary outer
codes. We show that some of the Varshamov-Tenengol'ts-Constantin-Rao
codes, a class of binary nonlinear codes for this channel, have a nice
structure when viewed as ternary codes.  In many cases, our ternary
construction yields even better codes. For the nonbinary asymmetric
channel, our methods construct linear codes for many lengths and
distances which are superior to the linear codes of the same length
capable of correcting the same number of symmetric errors.

In the binary case, Varshamov \cite{Var65} has shown that almost all
good linear codes for the asymmetric channel are also good for the
symmetric channel.  Our results indicate that Varshamov's argument
does not extend to the nonbinary case, i.e., one can find better
linear codes for asymmetric channels than for symmetric ones.
\end{abstract}

\section{Introduction}
\label{sec:intro}

In communication systems, the signal transmitted is conventionally
represented as a finite sequence of elements from an alphabet $A$,
which we assume to be finite.  In general, we may take
$A=\{0,1,\ldots,q-1\}$, and if needed, some additional structure is
assumed, e.g., $A=\mathbb{Z}_q$ or $A=\mathbb{F}_q$. The most commonly
discussed channel model is the uniform symmetric channel, that is, an
error $a\rightarrow b$ happens with equal probability for any $a,b\in
A$ and $a\neq b$.  Error-correcting codes for these channels are
extensively studied, see, for instance, \cite{MacWilliams}.

However, in other systems, such as some data storing systems including
flash memories \cite{CR79:2,flash} and optical communication
\cite{MR}, the probability of the error $a\rightarrow b$ is no longer
independent of $a$ and $b$ and might vary a lot.  If some errors of
low probability are neglected, some of those channels can be modeled
as `asymmetric channels.'

More precisely, let the alphabet be $A=\{0,1,\ldots,q-1\}\subset
\mathbb{Z}$ with the ordering $0<1<2<\cdots < q-1$. A channel is called
asymmetric if any transmitted symbol $a$ is received as $b\leq a$. For
example, for $q=2$, the symbol $0$ is always received correctly while
$1$ may be received as $0$ or $1$. The corresponding channel is
called $\mathcal{Z}$-channel, see Fig.~\ref{fig:channel}. For $q>2$,
one can have different types of asymmetric channels~\cite{DetCode}.

Coding problems for asymmetric channels were discussed by Varshamov in
1965 \cite{Var65}. For the characterization of codes for these
channels, we need the following.
\begin{definition}[see \cite{Var65,Var73,Klo81}]
\label{def:channelmodel}
For $\mathbf{x},\mathbf{y}\in A^n$, where
$\mathbf{x}=(x_1,x_2,\ldots,x_n)$ and
$\mathbf{y}=(y_1,y_2,\ldots,y_n)$, let
\begin{itemize}
\item[(i)] $w(\mathbf{x}):=\sum_{i=1}^n x_i$,
\item[(ii)]
  $N(\mathbf{x},\mathbf{y}):=\sum_{i=1}^{n}\max\{y_i-x_i,0\}$, and
\item[(iii)]
  $\Delta(\mathbf{x},\mathbf{y}):=\max\{N(\mathbf{x},\mathbf{y}),N(\mathbf{y},\mathbf{x})\}$. 
\end{itemize}
Here $w(\mathbf{x})$ is the weight of $\mathbf{x}$, and
$\Delta(\mathbf{x},\mathbf{y})$ is called the asymmetric distance
between $\mathbf{x}$ and $\mathbf{y}$.  If $\mathbf{x}$ is sent and
$\mathbf{y}$ is received, we say that $w(\mathbf{x}-\mathbf{y})$
errors have occurred.  Note that $w(\mathbf{x}-\mathbf{y})\ge 0$ for
asymmetric channels.
\end{definition}

In this model, a code correcting $t$-errors is called a
$t$-code~\cite{Klo81}. The following theorem naturally follows.
\begin{theorem}[see \cite{Klo81}]
\label{thm:main}
A set $\mathcal{C}\subset A^n$ is a $t$-code if and only if
$\Delta(\mathbf{x},\mathbf{y})>t$ for all $\mathbf{x},\mathbf{y}\in
\mathcal{C}$, $\mathbf{x}\neq\mathbf{y}$.
\end{theorem}

Apparently, any code which can correct $t$ errors on a symmetric
channel will also be capable of correcting $t$ asymmetric errors, but
the converse is not true in general.  However, Varshamov showed that
almost all linear binary codes which are able to correct $t$ errors
for the $\mathcal{Z}$-channel are also able to correct $t$ symmetric
errors \cite{Var65}.  Therefore, in order to construct good codes for
the $\mathcal{Z}$-channel, nonlinear constructions are needed.
Varshamov and Tenengol'ts \cite{Varshamov2}, followed by Constantin
and Rao \cite{constantin-rao}, constructed families of $1$-codes for
the $\mathcal{Z}$-channel with size $\geq\frac{2^n}{n+1}$.  These
codes are constructed based on an Abelian group $G$ for which the
group operation is denoted by `$+$' and the identity of $G$ is denoted
by $0_G$ or just $0$.
\begin{definition}[Constantin-Rao (CR) codes]
\label{def:CR}
Let $G$ be an Abelian group of order $n+1$ and identity $0_G$. For
fixed $g\in G$, the CR code $\mathcal{C}_g$ is given by
\begin{equation}\label{eq:CR}
\mathcal{C}_g=(\{(x_1,x_2,\ldots,x_n)|\sum_{i=1}^n x_ig_i=g\}),
\end{equation}
where $g_1,g_2,\ldots,g_n$ are the non-identity elements of $G$,
$x_i\in\{0,1\}$, and the product $x_i g_i$ is defined in the canonical
way $1 g_i=g_i$ and $0 g_i=0_G$.
\end{definition}

If the group $G$ is a cyclic group of order $n+1$, then the
corresponding codes are Varshamov-Tenengol'ts (VT) codes
\cite{Varshamov2} (denoted by $\mathcal{V}_g$).  It is known that the
largest Constantin-Rao code of length $n$ is the code $\mathcal{C}_0$
based on the group
$G=\bigoplus_{p|(n+1)}\bigoplus_{i=1}^{n_p}\mathbb{Z}_p$, where
$n+1=\Pi_{p|(n+1)}p^{n_p}$ is the prime factorization of $n+1$ and
$\oplus$ denotes the direct product of groups (see \cite{Klo81}).
These VT-CR codes have better rates than the corresponding
single-error-correcting codes for the binary symmetric channel for all
lengths $n$ apart from $n=2^r-1$. In this case, the code $\mathcal{C}_0$ for the
group $G=\mathbb{Z}_2^r$ is the linear binary Hamming code.

These VT-CR codes have a direct generalization to the nonbinary case.
The modification of Definition \ref{def:CR} is to let $x_i\in
A=\{0,1,\ldots,q-1\}$ and require that the order of $g_i$ is at least
$q$. The resulting nonlinear codes have cardinality
$|\mathcal{C}_g|\geq \frac{q^n}{n+1}$.  Note that by the Hamming
bound, we have $|\mathcal{C}_{\text{sym}}|\le\frac{q^n}{(q-1)n+1}$ for
a symmetric single-error-correcting code.  Hence for $q>2$ and all
lengths $n$, the VT-CR codes have more codewords than the best
single-error-correcting symmetric codes of the same length.  The
construction can also be generalized to the case of $t$-codes with
$t>1$, for both binary and nonbinary alphabets \cite{Klo81}.

Some other constructions for designing single-error-cor\-rect\-ing
codes for the $\mathcal{Z}$-channel have also been introduced.  In
particular the partition method,
together with some heuristic search give good lower bounds for small
length codes with $n\leq 25$ \cite{Etzion,Bassam,Etzion2,Shilo}.
Nevertheless, the VT-CR construction remains the best systematic
construction of binary $1$-codes to date, and the situation is similar
for the nonbinary case. For a survey of classical results on codes for
the $\mathcal{Z}$-channel, see \cite{Klo81}.

In this paper, we present new constructions of codes for asymmetric
channels for both binary and nonbinary alphabets, based on methods of
generalized code concatenation.  For the binary asymmetric channel,
our methods construct nonlinear $1$-codes from ternary outer codes
which are better than the VT-CR codes.  For nonbinary asymmetric
channels, our methods yield linear codes for many lengths and
distances, which outperform the linear codes of the same lengths
capable of correcting the same number of symmetric errors.  For
certain lengths, our construction gives linear codes with equal
cardinality as the nonlinear VT-CR codes.  
Our results indicate that Varshamov's argument
does not extend to the nonbinary case, i.e., one can find better
linear codes for asymmetric channels than for symmetric ones.
We will also apply our nonbinary linear codes to correct asymmetric
limited magnitude errors~\cite{CSB+10},  
which models the asymmetric errors in multilevel flash memories 
in a more detailed manner.

\section{Binary asymmetric codes from ternary outer codes}
\label{sec:ternary}
To discuss our new construction for asymmetric codes based on the
generalized concatenation method, we start with the binary case,
building $1$-codes for the $\cal{Z}$-channel. 
We know that in this case, good codes would have to be nonlinear, 
so our method returns nonlinear codes.  

To construct $1$-codes for the $\cal{Z}$-channel, we first partition
all two-bit strings $\{00,01,10,11\}$ into three $1$-codes, which are
$C_{\mathit{0}}=\{00,11\}$, $C_{\mathit{1}}=\{01\}$,
$C_{\mathit{2}}=\{10\}$. Then we further find some outer codes over
the alphabet $\{\mathit{0},\mathit{1},\mathit{2}\}$ (i.e. ternary
outer codes). Each code symbol is encoded into each of the $1$-codes
by $i\mapsto C_i$.  To be more precise, define a binary to ternary map
$\tilde{\mathfrak{S}}$, which maps two bits to one trit.
\begin{definition}
\label{def:Stilde}
The map $\tilde{\mathfrak{S}}\colon\: \mathbb{F}^2_2\rightarrow
\mathbb{F}_3$ is defined by
\begin{equation}
\tilde{\mathfrak{S}}\colon\: 00\mapsto\mathit{0},\ 11\mapsto\mathit{0},\ 01\mapsto\mathit{1},\ 10\mapsto\mathit{2}.
\end{equation}
\end{definition}

The encoding $i\rightarrow C_i$ is then given by the inverse map of
$\tilde{\mathfrak{S}}$.  Note that $\tilde{\mathfrak{S}}$ is not
one-to-one.  So for the ternary symbol $\mathit{0}$ the inverse map
gives the two binary codewords $00$ and $11$, while for $\mathit{1}$
and $\mathit{2}$ we get the unique codewords $01$ and $10$,
respectively.
\begin{definition}
\label{def:S}
The map $\mathfrak{S}\colon\: \mathbb{F}_3\rightarrow
\powerset(\mathbb{F}^2_2)$ is defined by
\begin{equation}
\mathfrak{S}\colon\: \mathit{0}\mapsto \{00,11\},\ \mathit{1}\mapsto\{01\},\ \mathit{2}\mapsto\{10\}.
\end{equation}
\end{definition}

Note that for a binary code of length $n=2m$, by choosing a pairing of
coordinates, the map $\tilde{\mathfrak{S}}^m\colon\:
\mathbb{F}^{2m}_{2}\rightarrow \mathbb{F}_3^m$ takes a given binary
code of length $2m$ to a ternary code of length $m$.  On the other
hand, Definition \ref{def:S} can be naturally extended as well, i.e.,
the map $\mathfrak{S}^m$ takes a given ternary code of length $m$ to a
binary code of length $2m$.  The map $\mathfrak{S}^m$ hence specifies
the encoding of an outer ternary code into the inner codes $C_i$.

We remark that our method 
is indeed a two-level concatenation as discussed in~\cite{Dumer}.
In the language of~\cite{Dumer}, we have an inner code $B_0=\{00,01,10,11\}$
which is partitioned into three codes $B_{1,1}=\{00,11\}$, $B_{1,2}=\{01\}$
and $B_{1,3}=\{10\}$. We also have two outer codes, one is a ternary code $A_0$
of length $m$, and the other is the trivial ternary code of length 1, i.e. $A_1=\{0,1,2\}$.
The two-level concatenated code is then a binary code with length $2m$.

For a better understanding of the maps $\tilde{\mathfrak{S}}^m$ and
$\mathfrak{S}^m$, we look at some examples.

\begin{example}
The optimal $1$-code $\mathcal{C}^{(4)}$ of length $n=4$ and
cardinality $4$ has four codewords $0000,1100,0011,1111$. By pairing
coordinates $1,2$ and $3,4$, the ternary image under
$\tilde{\mathfrak{S}}^2$ is then $\mathit{00}$.
\end{example}

\begin{example}
\label{Eg612}
By starting from the ternary outer code of length $n=3$ with the
codewords
$\mathit{000,111,122,212,221}$,
the map $\mathfrak{S}^3$ yields the binary code $\mathcal{C}^{(6)}$
with the $12$ codewords
\begin{equation}
\begin{array}{llllll}
000000, & 000011, & 001100, & 001111, & 110000, & 110011,\\
111100, & 111111, & 010101, & 011010, & 100110, & 101001.
\end{array}
\end{equation}
The code $\mathcal{C}^{(6)}$ has asymmetric distance $2$, hence
correcting one asymmetric error. This is known to be an optimal
$1$-code for $n=6$ \cite{Klo81}.
\end{example}

\begin{example}
\label{Eg832}
By starting from the linear ternary code $[4,2,3]_3$ with generators
$\mathit{0111,1012}$, the map $\mathfrak{S}^4$ yields the binary
code $\mathcal{C}^{(8)}$ with $32$ codewords
\begin{equation}
\begin{array}{llll}
00000000, & 00000011, & 00001100, & 00001111, \\
00110000, & 00110011, & 00111100, & 00111111, \\
11000000, & 11000011, & 11001100, & 11001111, \\
11110000, & 11110011, & 11111100, & 11111111, \\
00010101, & 00101010, & 11010101, & 11101010, \\
01000110, & 10001001, & 01110110, & 10111001, \\
01011000, & 10100100, & 01011011, & 10100111, \\
10010001, & 01100010, & 10011101, & 01101110.
\end{array}
\end{equation}
$\mathcal{C}^{(8)}$ has asymmetric distance $2$, hence
correcting one asymmetric error.  We observe that 
$\mathcal{C}^{(8)}$ is exactly the CR code
$\mathcal{C}_0$ of length $n=8$ constructed from the group
$\mathbb{Z}_3\oplus\mathbb{Z}_3$, which hints some relationship
between the ternary construction and CR codes. We will discuss
this in more detail in Sec. \ref{sec:Crternary}.
\end{example}

\begin{figure}[h!]
\centerline{\small%
\begin{tabular}{ccc}
&
\begin{picture}(80,20)(0,5)
\put(10,10){\makebox(0,0){$01$}}
\put(15,10){\vector(2,-1){20}}
\put(35,20){\vector(-2,-1){20}}
\put(40, 0){\makebox(0,0){$00$}}
\put(40,20){\makebox(0,0){$11$}}
\put(45,20){\vector(2,-1){20}}
\put(65,10){\vector(-2,-1){20}}
\put(70,10){\makebox(0,0){$10$}}
\end{picture}
&
\\
\begin{picture}(50,30)(0,5)
\put(10,10){\makebox(0,0){$0$}}
\put(40,10){\makebox(0,0){$1$}}
\put(37,10){\vector(-1,0){24}}
\end{picture}
&
\begin{picture}(80,30)(0,5)
\put(10,10){\makebox(0,0){$\mathit{1}$}}
\put(13,10){\vector(1,0){24}}
\put(37,10){\vector(-1,0){24}}
\put(40,10){\makebox(0,0){$\mathit{0}$}}
\put(43,10){\vector(1,0){24}}
\put(67,10){\vector(-1,0){24}}
\put(70,10){\makebox(0,0){$\mathit{2}$}}
\end{picture}
&
\begin{picture}(50,30)(0,5)
\put(25,33){\makebox(0,0){$\mathit{0}$}}
\put(10,10){\makebox(0,0){$\mathit{1}$}}
\put(40,10){\makebox(0,0){$\mathit{2}$}}
\put(37,10){\vector(-1,0){24}}
\put(13,10){\vector(1,0){24}}
\put(13,10){\vector(2,3){12}}
\put(25,28){\vector(-2,-3){12}}
\put(37,10){\vector(-2,3){12}}
\put(25,28){\vector(2,-3){12}}
\end{picture}
\\
$\mathcal{Z}$
&
$\mathcal{T}$
&
$\mathcal{R}_3$
\end{tabular}}
\caption{The binary asymmetric channel $\mathcal{Z}$, the ternary
  channel $\mathcal{T}$ derived from $\mathcal{Z}\times\mathcal{Z}$
  and $\tilde{\mathfrak{S}}$, and the ternary symmetric channel
  $\mathcal{R}_3$.  The arrows indicate the possible transitions
  between symbols.}
\label{fig:channel}
\end{figure}
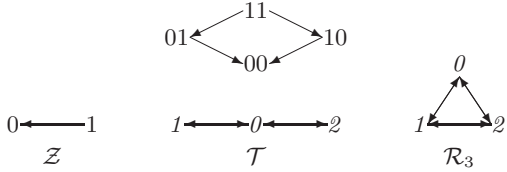

Example \ref{Eg832} indicates that good $1$-codes can be obtained from
some ternary codes under the map $\mathfrak{S}^m$.  Now the question
is what is the general condition under which a ternary code gives a
$1$-code via the map $\mathfrak{S}^m$. To address this question, by
combining the action of the channel $\mathcal{Z}\times\mathcal{Z}$ and
the map $\tilde{\mathfrak{S}}$, we obtain the ternary channel
$\mathcal{T}$ as shown in middle of Fig.~\ref{fig:channel}. Note that
$\mathcal{T}$ is different from the ternary symmetric channel
$\mathcal{R}_3$, which is also shown in Fig.~\ref{fig:channel}.

Now we come to the main result of this section,
which states that any single-error-correcting code for the ternary
channel $\mathcal{T}$ gives a $1$-code under the map $\mathfrak{S}^m$.
\begin{theorem}
If $\mathcal{C}'$ is a single-error-correcting ternary code of length
$m$ for the channel $\mathcal{T}$, then
$\mathcal{C}=\mathfrak{S}^{m}(\mathcal{C}')$ is a $1$-code of length
$2m$.
\label{ChannelT}
\end{theorem}
\begin{IEEEproof}
For any two codewords $\mathbf{c}'_1,\mathbf{c}'_2\in
\mathcal{C}'$, we have to show that the asymmetric distance between
$\mathfrak{S}^m(\mathbf{c}'_1)$ and $\mathfrak{S}^m(\mathbf{c}'_2)$ is
at least two.

First assume that the Hamming distance between $\mathbf{c}'_1$ and
$\mathbf{c}'_2$ is at least three. Then the Hamming distance between
$\mathfrak{S}^m(\mathbf{c}'_1)$ and $\mathfrak{S}^m(\mathbf{c}'_2)$ is
also at least three, which implies that the asymmetric distance between
$\mathfrak{S}^m(\mathbf{c}'_1)$ and $\mathfrak{S}^m(\mathbf{c}'_2)$ is
at least two.

If the Hamming distance between $\mathbf{c}'_1$ and $\mathbf{c}'_2$ is
less than three, it suffices to consider ternary words of length two.
It turns out that the following ten pairs of such ternary words can be
uniquely decoded if a single error happens in the channel
$\mathcal{T}\times\mathcal{T}$:
\begin{equation}
\begin{array}{lllll}
\mathit{01,22} & \mathit{10,22} & \mathit{01,12} & \mathit{10,21} & \mathit{02,11},\\
\mathit{20,11} & \mathit{02,21} & \mathit{20,12} & \mathit{11,22} & \mathit{12,21}.
\end{array}
\label{pairs}
\end{equation}
The asymmetric distance between the images of each pair under
$\mathfrak{S}^2$ is at least two.
\end{IEEEproof}

The following corollary is straightforward.
\begin{corollary}
\label{coro:d3}
If $\mathcal{C}'$ is an $(m,K,3)_3$ code, then
$\mathfrak{S}^{m}(\mathcal{C}')$ is a $1$-code of length $2m$.
\label{linear}
\end{corollary}

The size of the binary code can be computed as follows.
\begin{theorem}
Let $\mathcal{C}'$ be a ternary code of length $m$ with homogeneous
weight enumerator
\begin{equation}
W_{\mathcal{C}'}(X,Y)=\sum_{\mathbf{c}'\in\mathcal{C}'} X^{m-\wgt(\mathbf{c}')}Y^{\wgt(\mathbf{c}')},
\end{equation}
where $\wgt(\mathbf{c}')$ denotes the Hamming weight of
$\mathbf{c}'$. Then $\mathcal{C}=\mathfrak{S}^m(\mathcal{C}')$ has
cardinality $|\mathcal{C}|=W_{\mathcal{C}'}(2,1)$.
\end{theorem}
\begin{IEEEproof}
By Definition \ref{def:S}, for every zero in the codeword
$\mathbf{c}'$ the corresponding pair in the binary codeword can take
two different values, while the non-zero elements are mapped to a
unique binary string. Hence
$|\mathfrak{S}^m(\mathbf{c}')|=2^{m-\wgt(\mathbf{c}')}$.
\end{IEEEproof}

Theorem \ref{ChannelT} only works for designing $1$-codes of even
length.  So we generalize this construction to odd length, starting
from `adding a bit' to the ternary code.
\begin{theorem}
\label{th:odd}
If $\mathcal{C}'$ is a single-error-correcting code of length $m+1$
for the channel $\mathcal{Z}\times\mathcal{T}^{m}$, then
$\mathcal{C}=\mathfrak{S}^{m}(\mathcal{C}')$ is a $1$-code of length
$2m+1$, where $\mathfrak{S}^{m}$ acts on the last $m$ coordinates of
$\mathcal{C}'$.
\label{ChannelZT}
\end{theorem}
\begin{IEEEproof}
First note that the combined channel
$\mathcal{Z}\times\mathcal{T}^{m}$ has a mixed input alphabet. Hence
the first coordinate in $\mathcal{C}$ is binary while the others are
ternary. For any two codewords $\mathbf{c}'_1,\mathbf{c}'_2\in
\mathcal{C}'$, we have to show that the asymmetric distance between
$\mathfrak{S}^m(\mathbf{c}'_1)$ and $\mathfrak{S}^m(\mathbf{c}'_2)$ is
at least two.

First assume that the Hamming distance between $\mathbf{c}'_1$ and
$\mathbf{c}'_2$ is at least three. Then the Hamming distance between
$\mathfrak{S}^m(\mathbf{c}'_1)$ and $\mathfrak{S}^m(\mathbf{c}'_2)$ is
also at least three, implying that the asymmetric distance between
$\mathfrak{S}^m(\mathbf{c}'_1)$ and $\mathfrak{S}^m(\mathbf{c}'_2)$ is
at least two.

If the Hamming distance between $\mathbf{c}'_1$ and $\mathbf{c}'_2$ is
less than three, the case that the positions where they differ does
not involve the first coordinate has already been covered in the proof
of Theorem \ref{ChannelT}.  So assume that the first coordinate is a
bit and the second is a trit. There are exactly two pairs
$0\mathit{1},1\mathit{2}$ and $1\mathit{2},1\mathit{1}$ for which a
single error on $\mathcal{Z}\times\mathcal{T}$ can be corrected.  The
corresponding images of each pair under $\mathfrak{S}^m$ give binary
codewords of asymmetric distance two.
\end{IEEEproof}

To illustrate this construction for odd length codes, we look at the
following example.
\begin{example}
Consider the code $0\mathit{000}$, $0\mathit{111}$, $0\mathit{222}$,
$1\mathit{012}$, $1\mathit{120}$, $1\mathit{201}$ for the channel
$\mathcal{Z}\times\mathcal{T}^{3}$.  The image under the map
$\mathfrak{S}^3$ is the binary code
\begin{equation}
\begin{array}{llll}
0000000, & 0000011, & 0001100, & 0001111, \\
0110000, & 0110011, & 0111100, & 0111111, \\
0010101, & 0101010, & 1000110, & 1110110, \\
1011000, & 1011011, & 1100001, & 1101101. \\
\end{array}
\end{equation}
This is a code of length $7$, cardinality $16$, with asymmetric
distance two, hence correcting one asymmetric error.
\label{gt7}
\end{example}

The following corollary is straightforward, but gives the most general situation of the ternary construction.
\begin{corollary}
If $\mathcal{C}'$ is a ternary single error correcting code of channel $\mathcal{Z}^{{m_1}}\times\mathcal{T}^{{m_2}}$ of length $m_1+m_2$, then $\mathcal{C}=\mathfrak{S}^{m_2}(\mathcal{C}')$ is a $1$-code of length $m_1+2m_2$, where $\mathfrak{S}^{m_2}$ acts on the last $n$ coordinate of $\mathcal{C}'$.
\label{general}
\end{corollary}

\section{New binary asymmetric codes with structure}
\label{sec:structure}
\label{sec:linear}
In the following, we compare nonlinear binary codes for the
$\mathcal{Z}$-channel which are the image of ternary linear codes
(``$\mathbb{F}_3$-linear codes''), and linear binary codes. For this,
we compare the rate of $1$-codes for various length.  The ratio of the
rates is given by $s=\log_2|T|/\log_2|B|$, where $|T|$ and $|B|$ are
the cardinalities of the nonlinear binary $1$-code from a linear
ternary code $T$ of Hamming distance three, and a linear binary code
$B$ of Hamming distance three, respectively.
\begin{table}
\caption{Ratio $s$ of the rates of $\mathbb{F}_3$-linear codes
  and linear binary codes\label{tab:TableI}}
\centerline{$
\begin{array}{|c|c|c|c|c|c|c|c|c|}
\hline
n & 6     & 8     & 10    & 12    & 14    & 16    & 18    \\\hline 
s & 1.107 & 1.250 & 1.000 & 0.940 & 0.936 & 1.026 & 1.020 \\\hline 
\hline
n & 20    & 22    & 24    & 26    & 28    & 30    & 32    \\\hline 
s & 1.017 & 1.014 & 1.013 & 1.012 & 0.967 & 0.946 & 0.987 \\\hline 
\hline
n & 34    & 36    & 38    & 40    & 42    & 44    & 46    \\\hline 
s & 0.988 & 0.988 & 0.989 & 0.990 & 0.990 & 0.991 & 0.991 \\\hline 
\hline
n & 48    & 50    & 52    & 54   & 56    & 58    & 60     \\\hline 
s & 0.992 & 0.992 & 0.992 &0.993 & 0.993 & 0.993 & 0.994  \\\hline 
\hline
n & 62    & 64    & 66    & 68    & 70    & 72    & 74    \\\hline 
s & 0.994 & 1.012 & 1.011 & 1.011 & 1.010 & 1.010 & 1.010 \\\hline 
\hline
n & 76    & 78    & 80    & 82    & 84    & 86    & 88    \\\hline 
s & 1.010 & 1.009 & 1.009 & 0.987 & 0.988 & 0.988 & 0.988 \\\hline 
\end{array}
$}
\end{table}
\ignore{
for n:=3 to 128 do 2*n,Log(2,Evaluate(WeightEnumerator(BestDimensionLinearCode(GF(3),n,3)),[2,1]))/Dimension(BDLC(GF(2),2*n,3));end for;      
}

From Table \ref{tab:TableI} we see that for certain lengths, the
$1$-codes obtained from ternary linear codes indeed encode more bits
than the corresponding linear binary codes.  In particular, for $n=8$
the $1$-code of cardinality $32$ encodes one bit more than the linear
binary code of size $16$.  This should be related to the fact that the
ternary Hamming code of length $8/2=4$ is `good.' On the other hand,
binary linear codes of distance three are `bad' for length
$8$, $16$, $32$, $64$.
Also, the $1$-codes of length $64$ through $80$ outperform the
corresponding linear binary code, i.e. $s>1$.  A general understanding
of the condition under which $s>1$ for those $\mathbb{F}_3$-linear
codes for the $\mathcal{Z}$-channel is still lacking.  For instance,
we do not know why $s<1$ for $n=32$, despite the fact that the binary
linear code of distance three is `bad' at length $32$.

\label{sec:cyclic}

Recall that Example \ref{Eg612} starts from a single-error-correcting
ternary cyclic code of length $3$, and results in a $1$-code of length
$6$ achieving the upper bound given in \cite{Klo81} via the map
$\mathfrak{S}^3$. Note that by the ternary construction, ternary
cyclic codes give binary quasi-cyclic codes. 
It turns out that we can find more good $1$-codes from cyclic ternary
codes of length $m$.

For $m=4$, we have found a ternary cyclic code with codewords
$\mathit{0000}$, $\mathit{0112}$, $\mathit{1222}$, $\mathit{1111}$,
and their cyclic shifts, which leads to a $1$-code with parameters
$(8,29)$.  For $m=5$, we have found a unique ternary cyclic code which
lead to a $1$-code with parameters $(10,98)$.  For $m=6, 7, 8$, we
have found ternary cyclic codes which lead to $1$-codes with
parameters $(12,336)$, $(14,1200)$, and $(16,3952)$, respectively. 
The generators of the cyclic codes for $m=4,\ldots,8$ are given in
Table~\ref{tab:ternary_cyclic}.

\begin{table}
\caption{Generators of ternary cyclic codes which yield good
  binary $1$-codes.\label{tab:ternary_cyclic}}
\begin{center}
\begin{tabular}{|l|p{0.8\hsize}|}
\hline
$m$ & generators\\
\hline
$4$ & $\mathit{0000}$, $\mathit{0112}$, $\mathit{1222}$,
$\mathit{1111}$\\
\hline
$5$ &$\mathit{00000}$, $\mathit{10012}$, $\mathit{20110}$,
$\mathit{12210}$, $\mathit{11202}$, $\mathit{11111}$,
$\mathit{22122}$\\
\hline
$6$ &
$\mathit{000000}$, $\mathit{100021}$, $\mathit{122000}$,
$\mathit{010101}$, $\mathit{120102}$, $\mathit{101101}$,\newline
$\mathit{201102}$, $\mathit{101202}$, $\mathit{102012}$,
$\mathit{222102}$, $\mathit{202020}$, $\mathit{112011}$,\newline
$\mathit{220220}$
\\
\hline
$7$ &
$\mathit{0000000}$, $\mathit{0000121}$, $\mathit{1100022}$,
$\mathit{0022020}$, $\mathit{1110100}$,\newline
$\mathit{1020100}$, $\mathit{1002001}$, $\mathit{0021021}$,
$\mathit{2001011}$, $\mathit{1200211}$,\newline
$\mathit{2021200}$, $\mathit{0201220}$, $\mathit{1022200}$,
$\mathit{1221010}$, $\mathit{1012020}$,\newline
$\mathit{1021201}$, $\mathit{1022121}$, $\mathit{2221020}$,
$\mathit{0112122}$, $\mathit{1111121}$,\newline
$\mathit{1112221}$, $\mathit{1122112}$, $\mathit{2121211}$,
$\mathit{2221212}$, $\mathit{2222222}$\\
\hline
$8$ &
$\mathit{00000201}$, $\mathit{00010112}$, $\mathit{00011010}$,
$\mathit{00021200}$, $\mathit{00101210}$, $\mathit{00110011}$,
$\mathit{00121111}$, $\mathit{00222110}$, $\mathit{01011102}$,
$\mathit{01212210}$, $\mathit{02021002}$, $\mathit{02112201}$,
$\mathit{02211101}$, $\mathit{02211210}$, $\mathit{02211222}$,
$\mathit{10001122}$, $\mathit{10010210}$, $\mathit{10122021}$,
$\mathit{10122111}$, $\mathit{10202002}$, $\mathit{11021220}$,
$\mathit{11100200}$, $\mathit{11111111}$, $\mathit{11111210}$,
$\mathit{11120002}$, $\mathit{11222011}$, $\mathit{12001200}$,
$\mathit{12100120}$, $\mathit{12102200}$, $\mathit{12111211}$,
$\mathit{12112022}$, $\mathit{12121212}$, $\mathit{20010200}$,
$\mathit{20102201}$, $\mathit{20121212}$, $\mathit{20210101}$,
$\mathit{20222011}$, $\mathit{20222200}$, $\mathit{21100210}$,
$\mathit{21120111}$, $\mathit{21120120}$, $\mathit{21200221}$,
$\mathit{21212110}$, $\mathit{22000012}$, $\mathit{22000100}$,
$\mathit{22020201}$, $\mathit{22022000}$, $\mathit{22101102}$,
$\mathit{22101222}$, $\mathit{22102210}$, $\mathit{22120110}$,
$\mathit{22221221}$, $\mathit{22222222}$\\
\hline
\end{tabular}
\end{center}
\end{table}

From Table \ref{tab:TableII} below we see that the $1$-codes from
cyclic ternary codes are not as good as the codes $(8,32)$ (given in
Example \ref{Eg832}) and $(10,105)$, $(12,351)$ which are obtained via
random numerical search based on the ternary construction.  However,
with growing length imposing the cyclic structure reduces the search
complexity. The codes $(14,1200)$ and $(16,3952)$ listed in Table
\ref{tab:TableII}, for example, are obtained from ternary cyclic
codes of length $m=7$ and $m=8$, respectively, while non-exhaustive
randomized search did not yield anything better as the search space is
too large.

For odd length, we can use the following construction of
\emph{extended} ternary codes.
\begin{lemma}\label{lemma:extended_ternary}
Let $\mathcal{C}'$ be a ternary code of length $m$ which can be
decomposed into two subcode $\mathcal{C}'_0$ and $\mathcal{C}'_1$ such
that each code $\mathcal{C}'_i$ can correct a single error for the
channel $\mathcal{T}$ and for any pair of codewords
$c'_0\in\mathcal{C}'_0$ and $c'_1\in\mathcal{C}'_1$, the distance with
respect to the channel $\mathcal{T}$ is at least two. Then the image
of $\mathcal{C}''=0\mathcal{C}'_0\cup 1\mathcal{C}'_0$ under $\mathfrak{S}^m$ is a
$1$-code of length $2m+1$ for the asymmetric binary channel.
\end{lemma}
\begin{IEEEproof}
We only have to consider codewords of $\mathcal{C}''$ which differ in
the first position, i.e., $c_0''=0c_0'$ and $c_1''=1c_1'$. If the
Hamming distance between $c_0'$ and $c_1'$ is only one, then without
loss of generality, we can assume $c_0'=\mathit{1}v$ and
$c_1'=\mathit{2}v$, as only the symbols $\mathit{1}$ and $\mathit{2}$
have distance two with respect to the channel $\mathcal{T}$. Then the
images of $c_i''$ under $\mathfrak{S}^m$ are
$c_0=001\mathfrak{S}^{m-1}(v)$ and $c_1=110\mathfrak{S}^{m-1}(v)$.
Similarly, if $c_0'$ and $c_1'$ differ in at least two positions,
the images of $c_i''$ under $\mathfrak{S}^m$ will have asymmetric
distance greater than one.
\end{IEEEproof}

Generators for \emph{extended cyclic codes} based on Lemma
\ref{lemma:extended_ternary} are given in Table~\ref{tab:extended_ternary_cyclic}.

\begin{table}
\caption{Generators of extended ternary cyclic codes which yield good
  binary $1$-codes.\label{tab:extended_ternary_cyclic}}
\begin{center}
\begin{tabular}{|l|p{0.7\hsize}|}
\hline
$m$ & generators\\
\hline
$3$ & $0\mathit{000}$, $0\mathit{111}$, $0\mathit{222}$,\newline $1\mathit{210}$\\
\hline
$4$ & 
$0\mathit{0000}$, $0\mathit{0221}$, $0\mathit{1211}$, $0\mathit{2222}$,\newline
$1\mathit{1010}$, $1\mathit{2020}$, $1\mathit{1220}$\\
\hline
$5$ &$0\mathit{00000}$, $0\mathit{10021}$, $0\mathit{12102}$,
     $0\mathit{20111}$, $0\mathit{22201}$,\newline
     $0\mathit{11111}$, $0\mathit{22222}$,\newline
   $1\mathit{02210}$, $1\mathit{01020}$, $1\mathit{01212}$\\
\hline
$6$ &$0\mathit{100021}$, $0\mathit{122000}$, $0\mathit{100100}$,
$0\mathit{200200}$, $0\mathit{010101}$,\newline
$0\mathit{222010}$, $0\mathit{110201}$, $0\mathit{101202}$,
$0\mathit{202020}$, $0\mathit{111111}$,\newline
$0\mathit{221211}$, $0\mathit{212211}$, $0\mathit{222222}$,\newline
$1\mathit{022100}$, $1\mathit{112000}$, $1\mathit{001002}$,
$1\mathit{120102}$, $1\mathit{101101}$,\newline
$1\mathit{012111}$, $1\mathit{102012}$, $1\mathit{220220}$,
$1\mathit{122202}$, $1\mathit{211112}$,\newline
$1\mathit{211222}$, $1\mathit{121212}$ \\
\hline
$7$ &

$0\mathit{1100002}$, $0\mathit{0200100}$, $0\mathit{1200010}$, $0\mathit{0202200}$,\newline
$0\mathit{0112200}$, $0\mathit{1002120}$, $0\mathit{1001011}$, $0\mathit{1210020}$,\newline
$0\mathit{1222100}$, $0\mathit{0022202}$, $0\mathit{1221200}$, $0\mathit{0101121}$,\newline
$0\mathit{0210201}$, $0\mathit{1102220}$, $0\mathit{1020111}$, $0\mathit{1012211}$,\newline
$0\mathit{2021210}$, $0\mathit{0122221}$, $0\mathit{1112021}$, $0\mathit{1202221}$,\newline
$0\mathit{1111111}$, $0\mathit{1122112}$, $0\mathit{2222222}$,\newline
$1\mathit{0221000}$, $1\mathit{0102000}$, $1\mathit{0001101}$, $1\mathit{2000120}$,\newline
$1\mathit{2101100}$, $1\mathit{1100120}$, $1\mathit{1002202}$, $1\mathit{1200220}$,\newline
$1\mathit{1200211}$, $1\mathit{0012112}$, $1\mathit{1021210}$, $1\mathit{2201022}$,\newline
$1\mathit{1110220}$, $1\mathit{0111211}$, $1\mathit{1212210}$, $1\mathit{0202122}$,\newline
$1\mathit{0211212}$, $1\mathit{2202212}$, $1\mathit{1221221}$\\ \hline
\end{tabular}
\end{center}
\end{table}
\begin{example}
For $m=3$, consider the cyclic codes $\mathcal{C}'_0=\{\mathit{000},
\mathit{111}, \mathit{222}\}$, and $\mathcal{C}'_1=\{\mathit{210},\mathit{021},\mathit{102}\}$.
The image of $0\mathcal{C}'_0\cup 1\mathcal{C}'_1$ under
$\mathfrak{S}^3$ is
\begin{alignat*}{10}
&0000000, 0000011, 0001100, 0001111,\\
&0110000, 0110011, 0111100, 0111111,\\
&0010101, 0101010,\\
&1100100, 1100111, 1001001, 1111001, 1010010, 1011110.
\end{alignat*}
\end{example}

We finally note that we use nonlinear cyclic codes. This makes it
more complicated to find a systematic generalized construction for
larger length.

\section{The binary VT-CR codes viewed as ternary codes}
\label{sec:Crternary}

In this section we clarify the relationship between the ternary
construction and the VT-CR codes, by showing that certain VT-CR codes
are a special case of the ternary construction. We start from the
following.
\begin{definition}
A binary code $\mathcal{C}$ of even length $n=2m$ is called
\emph{ternary} if
$\mathfrak{S}^{m}(\tilde{\mathfrak{S}}^m(\mathcal{C}))=\mathcal{C}$.
\end{definition}

Based on this definition, if a binary code $\mathcal{C}$ of even
length is ternary, then it can be constructed from some ternary code
via the map $\mathfrak{S}$. The following theorem shows that
certain VT-CR codes are a special case of asymmetric codes constructed
from some ternary codes.
\begin{theorem}
\label{th:ternary}
For $n$ even, the VT code $\mathcal{V}_g$ and the CR code
$\mathcal{C}_g$ are ternary for any $g$.
\end{theorem}
\begin{IEEEproof}
Let $\mathcal{C}=\mathcal{V}_g$ or $\mathcal{C}=\mathcal{C}_g$.  We
only need to prove that there exists a pairing of the coordinates of
$\mathcal{C}$ such that for any codeword $v\in \mathcal{C}$ the
following holds: if for a pair $\alpha$ of coordinates the code
symbols of $v$ are $00$, denoted by $v|_{\alpha}=00$, then there
exists another codeword $v'\in \mathcal{C}$ with $v'|_{\alpha}=11$ and
$v'|_{\bar{\alpha}}=v|_{\bar{\alpha}}$.  Here $\bar{\alpha}$ denotes
all coordinates except the pair $\alpha$.

Both the VT code $\mathcal{V}_g$ and the CR code $\mathcal{C}_g$ are
defined by a group $G$ of odd order $n+1$, and the coordinates of the
codewords correspond to the non-identity group elements. As the group
order is odd, the only group element that is its own inverse is
identity.  Hence we can pair every non-identity element $h\in G$ with
its inverse $-h$.  If neither $h$ nor $-h$ are contained in the sum in
Eq. (\ref{eq:CR}), then the sum clearly does not change when including
both $h$ and $-h$.
\end{IEEEproof}

We look at some examples.

\begin{example}
\label{eq:n6}
For $n=6$, the VT code $\mathcal{V}_0$ is given by
\begin{equation}
x_1+2x_2+3x_3+4x_4+5x_5+6x_6=0\bmod 7,
\end{equation}
where $x_i\in \{0,1\}$. Then one can use the pairing 
\begin{equation}
\{x_1x_6,x_2x_5,x_3x_4\}.
\end{equation}
The cardinality of the code is $10$.
The image of this code under $\tilde{\mathfrak{S}}^{6}$ is a linear code $[3,1,3]_3$. 
\end{example}

\begin{example}
For $n=8$, the VT code $\mathcal{V}_0$ is given by
\begin{equation}
\sum_{i=1}^8 ix_i=0\bmod 9,
\end{equation}
where $x_i\in \{0,1\}$. Then one can use the pairing 
\begin{equation}
\{x_1x_8,x_2x_7,x_3x_6,x_4x_5\}.
\end{equation}
The cardinality of the code is $30$.
\end{example}

\begin{example}
For $n=8$, the CR code $\mathcal{C}_0$ of largest cardinality, 
which is associated with the group $\mathbb{Z}_3\oplus \mathbb{Z}_3$, is given by 
\begin{alignat}{5}
x_1(0,1)+x_2(0,2)+x_3(1,0)+x_4(1,1)&\nonumber\\
+x_5(1,2)+x_6(2,0)+x_7(2,1)+x_8(2,2)&{}=\bmod (3,3),
\end{alignat}
where $x_i\in \{0,1\}$. Then one can use the pairing 
\begin{equation}
\{x_1x_2,x_3x_6,x_4x_8,x_5x_7\}.
\end{equation}
The cardinality of the code is $32$, which is however nonlinear.
The image of this code under $\tilde{\mathfrak{S}}^{8}$ is a linear
code $[4,2,3]_3$, which is the one given in Example \ref{Eg832}.
\end{example}

\begin{example}
Consider $n=10$. For the VT code $\mathcal{V}_0$ is then given by
\begin{equation}
\sum_{i=1}^{10}ix_i=0 \bmod 11,
\end{equation}
where $x_i\in \{0,1\}$. Then one can use the pairing 
\begin{equation}
\{x_1x_{10},x_2x_9,x_3x_8,x_4x_7,x_5x_6\}.
\end{equation}
The cardinality of the code is $94$, and the image of this code under
$\tilde{\mathfrak{S}}^{10}$ is equivalent to a cyclic ternary code
with $m=5$. Note that there exists a $1$-code $(10,98)$ which is
obtained from a cyclic ternary code (see Sec.~\ref{sec:cyclic}).
\end{example}

Now we consider the case of odd length.
\begin{definition}
A binary code $\mathcal{C}$ of odd length $n=2m+1$ is called
\emph{generalized ternary} if
$\mathfrak{S}^{m}(\tilde{\mathfrak{S}}^m(\mathcal{C}))=\mathcal{C}$,
where $\tilde{\mathfrak{S}}^m$ acts on the last $2m$ coordinates of
$\mathcal{C}$.
\end{definition}

Based on this definition, if a binary code $\mathcal{C}$ of odd length
$2m+1$ is generalized ternary, then it can be constructed from some
single-error-correcting code for the channel
$\mathcal{Z}\times\mathcal{T}^m$ via the map
$\tilde{\mathfrak{S}}$.
\begin{theorem}
\label{th:geternary}
For $n$ odd, the VT code $\mathcal{V}_g$ is generalized ternary for
any $g$. 
\end{theorem}
\begin{IEEEproof}
We only need to prove that there exists a pairing which leaves a
single coordinate as a bit, such that for any codeword $v\in
\mathcal{V}_g$, if $v$ restricted to a chosen pair $\alpha$ is $00$,
then there exist another codeword $v'\in \mathcal{V}_g$ such that
$v'|_{\alpha}=11$ and $v'|_{\tilde{\alpha}}=v|_{\tilde{\alpha}}$.

For a VT code $\mathcal{V}_g$ of odd length, choose the pairing
$\{i,n+1-i\}_{i=1}^{n/2}$ and leave the coordinate $(n+1)/2$ as a bit.
Then the above condition is satisfied.
\end{IEEEproof}

We discuss an example.

\begin{example}
For $n=7$, the VT code $\mathcal{V}_0$ is given by
\begin{equation}
\sum_{i=1}^7 ix_i=0 \bmod 8,
\end{equation}
where $x_i\in \{0,1\}$. Then one can use the pairing 
\begin{equation}
\{x_1x_7,x_2x_6,x_3x_5\}, 
\end{equation}
and treat $x_4$ as a bit. The size of the code is $16$, and it is
equivalent to the code given in Example \ref{gt7}.
\end{example}

\begin{table}[h!]
\begin{center}
\caption{Size of $1$-codes from ternary construction via numerical
  search, compared to CR codes, codes obtained by the partition
  method, and the known bounds from
  \cite{Klo81,Sloane:graphs,Etzion2}.\newline
  For odd $n$, `cyclic ternary' refers to extended cyclic codes.
  \label{tab:TableII}}
\begin{tabular}{cccccc}
\\
\hline
$n$ & CR & cyclic ternary & ternary & partition & known bounds\\
\hline
 6 &   10 &  12  &   12 &  *          & 12 \\
 7 &   16 &  16  &   16 &  *          & 18 \\
 8 &   32 &  29  &   32 &  *          & 36 \\
 9 &   52 &  53  &   55 &  *          & 62 \\
10 &   94 &  98  &  105 &  104$^{(a)}$ & 112--117 \\
11 &  172 & 154  &  180 &  180$^{(b)}$ & 198--210 \\
12 &  316 & 336  &  351 &  336$^{(b)}$ & 379--410 \\
13 &  586 & 612  &  612 &  652$^{(b)}$ & 699--786  \\
14 & 1096 & 1200 & 1200 & 1228$^{(b)}$ & 1273--1500 \\
15 & 2048 & 2144 & 2144 & 2288$^{(b)}$ & 2288--2828 \\
16 & 3856 & 3952 & 3952 & 4280$^{(b)}$ & 4280--5486 \\ 
\hline
\end{tabular}
\end{center}
\vskip-4ex
\end{table}
In Table \ref{tab:TableII}, the cardinality of codes found by the
(generalized) ternary method is compared to the size of the
corresponding VT-CR codes.  One can see that the
(generalized) ternary construction indeed outperforms the VT-CR
construction, in particular for larger $n$.

The column in Table \ref{tab:TableII} labeled `partition' is
obtained from the partition method in Ref. \cite{Etzion}. The code
$(a)$ is found from the partition of constant weight codes of length
$6$ and asymmetric codes of length $4$. Codes $(b)$
are from Ref. \cite{Etzion}.
For $n=10,11,12$, the ternary construction yields codes of equal size
or even more codewords compared to the partition method.  However,
the best codes are obtained by heuristic methods, which, e.g., give
$(10,112)$ \cite{Etzion2} and $(12,379)$ \cite{Shilo}. This is not
surprising as both the ternary construction and the partition
method assume some additional structure of the binary $1$-codes.

\section{Nonbinary asymmetric-error-correcting codes}
\label{sec:nonbinary}
In this section, we consider the construction of $1$-codes for
nonbinary asymmetric channels.  Recall that the characteristic
properties of codes for this channel model are given by
Definition~\ref{def:channelmodel} and Theorem \ref{thm:main}. Our
construction will again be based on concatenation, generalizing the
map $\mathfrak{S}^m$.

\begin{figure}[h!]
\centerline{\footnotesize\unitlength0.8\unitlength%
\begin{tabular}{ccc}
\begin{picture}(50,50)(0,5)
\put(25,33){\makebox(0,0){$0$}}
\put(10,10){\makebox(0,0){$1$}}
\put(40,10){\makebox(0,0){$2$}}
\put(37,10){\vector(-1,0){24}}
\put(13,10){\vector(1,0){24}}
\put(13,10){\vector(2,3){12}}
\put(25,28){\vector(-2,-3){12}}
\put(37,10){\vector(-2,3){12}}
\put(25,28){\vector(2,-3){12}}
\end{picture}
&\unitlength0.9\unitlength
\begin{picture}(50,50)(0,5)
\put(10,40){\makebox(0,0){$0$}}
\put(10,10){\makebox(0,0){$1$}}
\put(40,10){\makebox(0,0){$2$}}
\put(40,40){\makebox(0,0){$3$}}
\put(13,10){\vector(1,0){24}}
\put(37,10){\vector(-1,0){24}}
\put(13,40){\vector(1,0){24}}
\put(37,40){\vector(-1,0){24}}
\put(10,15){\vector(0,1){20}}
\put(10,35){\vector(0,-1){20}}
\put(40,15){\vector(0,1){20}}
\put(40,35){\vector(0,-1){20}}
\end{picture}
&\unitlength0.8\unitlength
\begin{picture}(50,50)(0,5)
\put(25,50){\makebox(0,0)[b]{$0$}}
\put(5,32.5){\makebox(0,0)[r]{$1$}}
\put(10,8){\makebox(0,0){$2$}}
\put(40,8){\makebox(0,0){$3$}}
\put(45,32.5){\makebox(0,0)[l]{$4$}}
\put(37,10){\vector(-1,0){24}}
\put(13,10){\vector(1,0){24}}
\put(13,10){\vector(-1,3){7.5}}
\put(5.5,32.5){\vector(1,-3){7.5}}
\put(37,10){\vector(1,3){7.5}}
\put(44.5,32.5){\vector(-1,-3){7.5}}
\put(5.5,32.5){\vector(4,3){19.5}}
\put(25,47.5){\vector(-4,-3){19.5}}
\put(25,47.5){\vector(4,-3){19.5}}
\put(44.5,32.5){\vector(-4,3){19.5}}
\end{picture}
\\
$\mathcal{R}_3$
&
$\mathcal{R}_4$
&
$\mathcal{R}_5$
\end{tabular}}
\caption{The induced channel $\mathcal{R}_3$ for $q=3$ (which is just
  the ternary symmetric channel), the induced channel $\mathcal{R}_4$
  for $q=4$, and the induced channel $\mathcal{R}_5$ for $q=5$. The
  arrows indicate the possible transitions between symbols.}
\label{fig:channel2}
\end{figure}
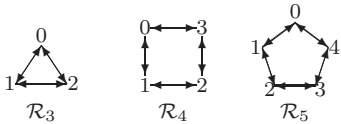

For a given $q$, choose the outer code as some code over the alphabet
$A=\{0, 1, \ldots,q-1\}$, which encodes to some inner codes
$\{C_0,C_1,\ldots,C_{q-1}\}$ via $i\mapsto C_i$.  Now choose the $q$
inner codes as the double-repetition code
$C_0=\{00,11,\ldots,(q-1)(q-1)\}$ and all its $q-1$ cosets
$C_i=C_0+(0i)$, i.e., we have the rule that $(0i)\in C_i$.  It is
straightforward to check that each $C_i$ is a $1$-code, i.e., has
asymmetric distance $2$.  Note that a single asymmetric error will
only drive transitions between $i,j$ for $i=j\pm 1$. For instance, for
$q=3,4,5$, the induced channels
$\mathcal{R}_3$, $\mathcal{R}_4$, $\mathcal{R}_5$ are shown in
Fig.~\ref{fig:channel2}. In general, we will write the induced channel
as $\mathcal{R}_q$ for outer codes over the alphabet $A=\{0, 1,
\ldots,q-1\}$.

Similar to Theorems~\ref{ChannelT} and \ref{th:odd}, we have the
following
\begin{theorem}
\label{th:nonbi1}
For $n$ even, an outer $(n/2,K)_q$ code correcting a single error for
the channel $\mathcal{R}_q$ leads to an $(n,q^{n/2}K)_q$ $1$-code
$\mathcal{C}$, for $q>2$. For $n$ odd, an outer $((n+1)/2,K)_q$ code
correcting a single error for the channel $\mathcal{R}_q$ leads to an
$(n,q^{(n-1)/2}K)_q$ $1$-code $\mathcal{C}$, for $q>2$.
\end{theorem}

If the outer code is linear, then our construction gives linear codes
for the asymmetric channel.
We state this result as a corollary below.
\begin{corollary}
\label{co:linear}
An outer $[m,k]_q$ linear code correcting a single error for the
channel $\mathcal{R}_q$ leads to a $[2m,m+k]_q$ $1$-code and a
$[2m-1,m+k-1]_q$ $1$-code, for $q>2$.
\end{corollary}

It turns out that in many cases, our construction gives linear codes
with larger cardinality than the distance-three symmetric codes of
equal length.  We first discuss the case of $q=3$.  In this case,
$\mathcal{R}_3$ is the ternary symmetric channel, so we will just use
outer codes of Hamming distance $3$. We consider some examples. 
\begin{example}
\label{eg:53}
Consider $q=3$ and take the outer code as $[3,1,3]_3$, with codewords
$\mathit{000}$, $\mathit{111}$, $\mathit{222}$.  This will give a
$[5,3]_3$ $1$-code
with codewords 
\begin{equation}
\begin{array}{lllll}
00000, & 00011, & 00022, & 01100, & 01111,\\ 
01122, & 02200, & 02211, & 02222, & \ \\
10101, & 10112, & 10120, & 11201, & 11212, \\
11220, & 12001, & 12012, & 12020, & \ \\
21010, & 21021, & 21002, & 22110, & 22121,\\
22102, & 20210, & 20221, & 20202, & \ \\
\end{array}
\label{code:53q3}
\end{equation}
while the best linear
single-symmetric-error-correcting code is $[5,2,3]_3$.
The $[3,1,3]_3$ outer code also yields a $[6,4]_3$ $1$-code, 
while the best linear single-symmetric-error-correcting code is
$[6,3,3]_3$.
Now take the outer code as $[4,2,3]_3$. 
This will give a $[7,5]_3$ $1$-code, while the best linear
single-symmetric-error-correcting code  is $[7,4,3]_3$. 
We can also construct  a $[8,6]_3$ $1$-code, while the best linear
single-symmetric-error-correcting code is $[8,5,3]_3$.
\end{example}

This example can be directly generalized to other
lengths. Furthermore, the constructions extend trivially to $q>3$, as
any code of Hamming distance $3$ corrects a single error for the
channel $\mathcal{R}_q$.  Note that Hamming codes over $\mathbb{F}_q$
have length $n_r=(q^r-1)/(q-1)$.  For a given $n_r$, our construction
then allows to construct asymmetric $1$-codes of all length
$[n_r+1,2n_r]$ for $n_r$ odd or all lengths $[n_r+2,2n_r]$ for $n_r$
even.  The sequence of lengths $n_r$ is a geometric series, and hence our method 
can construct asymmetric codes for approximately $\frac{1}{q}$ of all
lengths, outperforming the best single-symmetric-error-correcting linear codes.

Now consider the case $q>3$ in more detail.  The channel
$\mathcal{R}_q$ (see Fig. \ref{fig:channel2}) is no longer a symmetric
channel, so outer codes of Hamming distance $3$ are no longer expected
to give the best $1$-codes. It turns out, however, that
single-error-correcting codes for the channel $\mathcal{R}_q$ are
equivalent to single-symmetric-error correcting codes with respect to
Lee metric \cite{Ber68} (see also~\cite{KBE11}), for which optimal
linear codes are known.  When $q$ is odd, let $H$ be the parity check
matrix whose columns are all vectors in $\mathbb{Z}_q^r$ whose first non-zero
elements is in the $\{1,2,\ldots,\frac{q-1}{2}\}$ (where $r$ is the
number of rows in $H$), then the corresponding code can correct a
single error for the channel $\mathcal{R}_q$.

We consider an example.
\begin{example}
For $q=5$ consider the parity check matrix
\[
\left( 
\begin{array}{cccccccccc}
1 & 1 & 1 & 1 & 1 & 2 & 2 & 2 & 2 & 2 \\
0 & 1 & 2 & 3 & 4 & 0 & 1 & 2 & 3 & 4 
\end{array}
\right),
\]
which gives a $[10,8]_5$ code correcting a single error for the
channel $\mathcal{R}_5$, and hence a $[20,18]_5$ $1$-code. Note that
the best linear single-symmetric-error-correcting code for $n=20$ is
$[20,17,3]_5$.
\end{example}

Our new linear codes for asymmetric channels for $q>2$ show that
Varshamov's argument that for the binary case, there is almost no hope
to find good linear codes for the asymmetric channel, does not hold
for the nonbinary case. There is indeed room for constructing good
linear codes adapted to the asymmetric channel.

Note that contrary to the binary case, 
the nonlinear VT-CR codes can no longer be viewed as a special case
of our construction.
However, for lengths $n_r=q^r-1$, our construction gives codes of the
same cardinality as the VT-CR codes, while our codes are linear, but
the VT-CR codes are not. 

Finally, we briefly discuss the extension of our concatenation method
to construct $t$-asymmetric-error-correcting codes for $t>1$. We look
at some examples.
\begin{example}
Consider the case of $q=3$. Take the outer code as the $[5,3]_3$
$1$-code constructed in Example~\ref{eg:53}, which has asymmetric
distance $2$.  Now take the encoding to the inner code as
$\mathit{0\mapsto 00}$, $\mathit{1\mapsto 11}$, $\mathit{2\mapsto
  22}$.  Then the concatenated code has asymmetric distance $4$, which
gives a $[10,3]_3$ $3$-code, while the best linear
triple-error-correcting code is $[10,2,7]_3$.  Similarly, take the
outer code as the $[6,4]_3$ $1$-code, then the concatenated code is a
$[12,4]_3$ $3$-code, while the best $3$-error-correcting linear code
is $[12,3,7]_3$.
\end{example}

\section{Codes for asymmetric limited-magnitude errors}
\label{sec:limited}

In this section, we discuss the application of these nonbinary linear codes constructed in Sec. V to correct asymmetric limited-magnitude errors with wrap around. 
This new `asymmetric limited-magnitude error' model, is introduced recently in~\cite{CSB+10}, which models the asymmetric errors in multilevel flash memories in a more detailed manner. This model is parameterized by two integer parameters: $\tilde{t}$ is the maximum number of symbol errors within a codeword, and $\ell$ the maximal magnitude of an error. The definition of asymmetric limited-magnitude errors is the following~\cite{CSB+10}.
\begin{definition}
\label{def:limited}
A vector of integers $\mathbf{e}=(e_1,\ldots,e_i)$ is called a $\tilde{t}$ asymmetric $\ell$-limited-magnitude error word if $|\{i:e_i\neq 0|\leq\tilde{t}$,
and for all $i$, $0\leq e_i\leq\ell$.
\end{definition}

Here by `asymmetric' it still means that if any transmitted symbol $a$ is received as $b\leq a$. For a codeword $\mathbf{x}\in A^n$, 
then a $\tilde{t}$ asymmetric $\ell$-limited-magnitude channel outputs a vector $\mathbf{y}\in A^n$ such that
$\mathbf{y}=\mathbf{x}-\mathbf{e}$, where $\mathbf{e}$ is a $\tilde{t}$ asymmetric $\ell$-limited-magnitude error word.

Coding problems for these channels have an intimate relation to coding problems for asymmetric channels. Indeed, when $t=\tilde{t}\ell$, any $t$-code for the asymmetric channel trivially corrects $\tilde{t}$ asymmetric $\ell$-limited-magnitude errors. Of course, the reverse is not true. 

A generalization of Definition~\ref{def:limited} is when we allow asymmetric errors to wrap around from $0$ back to $q-1$.  That is, we interpret `$-$' in 
$\mathbf{y}=\mathbf{x}-\mathbf{e}$ as subtraction mod $q$.  This error model is then called
the asymmetric $\ell$-limited-magnitude channels with wrap around.

Similar as the asymmetric distance $\Delta(\mathbf{x},\mathbf{y})$, we can define a distance $d_{\ell}$ for this error model, as below.
\begin{definition}
\label{def:asydisell}
For $\mathbf{x},\mathbf{y}\in A^n$, define
$M(\mathbf{x},\mathbf{y})=|\{i:x_i>y_i\}|$. The distance $d_{\ell}$
between the words $\mathbf{x},\mathbf{y}$ is then defined as
\[
d_{\ell}(\mathbf{x},\mathbf{y})= \begin{cases}
n+1
&\text{if $\max_{i}\{|x_i-y_i|\}>\ell$}\\\max\{M(\mathbf{x},\mathbf{y}),M(\mathbf{y},\mathbf{x})\}&\text{otherwise}
\end{cases}
\]
\end{definition}

Similar as Theorem~\ref{thm:main}, the proposition below directly follows~\cite{CSB+10}.
\begin{proposition}
\label{pro:dell}
A code $\cal{C}$ corrects $\tilde{t}$ asymmetric $\ell$-limited-magnitude errors if and only
if $d_{\ell}(\mathbf{x},\mathbf{y})\geq\tilde{t}+1$ for all distinct $\mathbf{x},\mathbf{y}\in\cal{C}$.
\end{proposition}

And one can readily interpret $d_{\ell}$ for asymmetric $\ell$-limited-magnitude channels with wrap around (interpret `$-$' in as subtraction mod $q$),
such that Proposition~\ref{pro:dell} still holds. Apparently, in
general a $t$-code for the asymmetric channel can no longer be used to
correct errors for asymmetric $\ell$-limited-magnitude channel with
wrap around. There is a sphere packing bound which naturally follows.
\begin{theorem}~\cite{CSB+10}
\label{th:sphere}
If $\cal{C}$ is a $\tilde{t}$ asymmetric $\ell$-limited-magnitude
(with wrap-around) error-correcting code, of length $n$ over an
alphabet of size $q$, then 
\begin{equation}
\label{eq:sphere}
|\mathcal{C}|\sum_{i=0}^t{n\choose i}\ell^i\leq q^n.
\end{equation}
\end{theorem}
An asymmetric $\ell$-limited-magnitude code is called perfect in a
sense that it attains this sphere-packing bound.

Code designs for correcting asymmetric $\ell$-limited-mag\-ni\-tude
errors, with or without wrap around, are discussed
in~\cite{CSB+10,KBE11}. Here we show that the linear codes constructed
in Sec.~\ref{sec:nonbinary} can be used to correct asymmetric
$\ell$-limited-magnitude errors and then further discuss their
optimality using the sphere-packing bound.

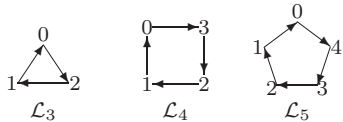
\begin{figure}[h!]
\centerline{\footnotesize\unitlength0.8\unitlength%
\begin{tabular}{ccc}
\begin{picture}(50,50)(0,5)
\put(25,33){\makebox(0,0){$0$}}
\put(10,10){\makebox(0,0){$1$}}
\put(40,10){\makebox(0,0){$2$}}
\put(37,10){\vector(-1,0){24}}
\put(13,10){\vector(2,3){12}}
\put(25,28){\vector(2,-3){12}}
\end{picture}
&\unitlength0.9\unitlength
\begin{picture}(50,50)(0,5)
\put(10,40){\makebox(0,0){$0$}}
\put(10,10){\makebox(0,0){$1$}}
\put(40,10){\makebox(0,0){$2$}}
\put(40,40){\makebox(0,0){$3$}}
\put(37,10){\vector(-1,0){24}}
\put(13,40){\vector(1,0){24}}
\put(10,15){\vector(0,1){20}}
\put(40,35){\vector(0,-1){20}}
\end{picture}
&\unitlength0.8\unitlength
\begin{picture}(50,50)(0,5)
\put(25,50){\makebox(0,0)[b]{$0$}}
\put(5,32.5){\makebox(0,0)[r]{$1$}}
\put(10,8){\makebox(0,0){$2$}}
\put(40,8){\makebox(0,0){$3$}}
\put(45,32.5){\makebox(0,0)[l]{$4$}}
\put(37,10){\vector(-1,0){24}}
\put(13,10){\vector(-1,3){7.5}}
\put(44.5,32.5){\vector(-1,-3){7.5}}
\put(5.5,32.5){\vector(4,3){19.5}}
\put(25,47.5){\vector(4,-3){19.5}}
\end{picture}
\\
$\mathcal{L}_3$
&
$\mathcal{L}_4$
&
$\mathcal{L}_5$
\end{tabular}}
\caption{ The asymmetric $1$-limited-magnitude channels $\mathcal{L}_n$
  for $n=3,4,5$ bits. The
  arrows indicate the possible transitions between symbols.}
\label{fig:channel3}
\end{figure}

Recall the construction in Sec.~\ref{sec:nonbinary}, where for a given
$q$, we choose the $q$ inner codes $C_0$, $C_1$, \ldots, $C_{q-1}$ as
$\{00,11,\ldots,{(q-1)(q-1)}\}$ and all its $q-1$ cosets.  It is
straightforward to check that each $C_i$ has $d_{\ell}=2$, for the
asymmetric $\ell$-limited-magnitude channel with wrap around, for
$\ell=1$, according to Definition~\ref{def:asydisell}.  Indeed, this
asymmetric $\ell$-limited-magnitude channel with wrap around, for
$\ell=1$ has transitions
\begin{equation}
(q-1)\rightarrow (q-2)\rightarrow (q-3)\cdots\rightarrow 1\rightarrow 0\rightarrow (q-1).
\end{equation}
We illustrate these asymmetric $1$-limited-magnitude channels $\mathcal{L}_n$
for $n=3,4,5$ bits in Fig.~\ref{fig:channel3}.

Now choose the outer code as some distance $3$ code over the alphabet $A=\{0, 1, \ldots,q-1\}$,
which encodes to the inner codes $\{C_0,C_1,\ldots,C_{q-1}\}$ via $i\rightarrow C_i$,
then the following results readily hold according to Proposition~\ref{pro:dell}.
\begin{proposition}
The codes based on the constructions given by Theorem~\ref{th:nonbi1}
and Corollary~\ref{co:linear} in Sec.~\ref{sec:nonbinary} correct a
single asymmetric $\ell$-limited-magnitude error with wrap around, for
$\ell=1$.
\end{proposition}
For $\ell=1$, the sphere-packing bound of Eq. (\ref{eq:sphere}) for
correcting a single error becomes
\begin{equation}
\label{eq:sphere1}
|\mathcal{C}|\leq\frac{q^n}{n+1}.
\end{equation}

Recall that for a given $n_r$, the construction in
Sec.~\ref{sec:nonbinary} gives linear codes of all lengths
$[n_r+1,2n_r]$ for $n_r$ odd or all lengths $[n_r+2,2n_r]$ for $n_r$
even, which outperform the best single-symmetric-error-correcting
linear codes.  Here $n_r=\frac{1}{q}(q^r-1)$, and $q$ is a prime power
that $\mathbb{F}_q$ is a field.  The sphere-packing bound given in
Eq. (\ref{eq:sphere1}) then shows that all these linear codes are
indeed optimal linear codes, for correcting a single asymmetric
$\ell$-limited-magnitude error with wrap around, for $\ell=1$.  For
$n=q^r-1$, we have perfect linear codes. As an example, the
$[8,6]_3$ code constructed in Example~\ref{eg:53} is a perfect linear
code. Note that perfect linear codes of length $n=q^r-1$ for
correcting a single asymmetric $\ell$-limited-magnitude error with
wrap around, for $\ell=1$, are also obtained in~\cite{KBE11}, but from
different constructions.

Indeed, those linear $t$-codes constructed in Sec.~\ref{sec:nonbinary} can also be used to correct
$\tilde{t}$ asymmetric $\ell$-limited-magnitude errors for $t=\tilde{t}\ell$. However, the sphere-packing
bound no longer tells us whether these linear codes are optimal.

\section{Discussion}
\label{sec:discussion}

We present new methods of constructing codes for asymmetric channels,
based on modified code concatenation. Our methods apply to both binary
and nonbinary case, for constructing both single- and
multi-asymmetric-error-correcting codes.

For the binary case, our construction gives nonlinear $1$-codes for
the $\cal{Z}$ channel, based on ternary outer codes. Some good
$1$-codes with structure, such as codes from ternary linear codes and
ternary cyclic codes are constructed. We also show that the VT-CR
code, which are the best known systematic construction of $1$-codes,
posses some nice structure while viewed in the (generalized) ternary
construction, and they are suboptimal under the (generalized) ternary
construction.  Indeed, this ternary construction is originally
inspired by constructing high performance quantum codes adapted to
asymmetric channels, see \cite{SSSZ09}.

For the nonbinary case, our construction gives linear $1$-codes, which
for many lengths outperforms the best
single-symmetric-error-correcting codes of the same lengths. Our
method can also be applied to construct good linear $t$-codes. To our
knowledge, our method gives the first systematic construction of good
linear codes for nonbinary asymmetric channels, which indicates that
Varshamov's argument of no good linear codes for asymmetric channels
does not extend to the nonbinary case.

Our $t$-codes also apply to correct $\tilde{t}$ asymmetric
$\ell$-limited-magnitude errors with wrap around, for
$t=\tilde{t}\ell$. These channels model the errors in multilevel flash
memory in a more detailed manner than Varshamov's asymmetric channel
given in Definition~\ref{def:channelmodel}. In case of $\ell=1$, our
single-error-correcting codes are shown to be optimal linear codes by
the sphere-packing bound. For lengths $n=q^r-1$, these codes are
perfect linear codes.

We hope our methods shade light on further study of asymmetric codes,
particularly, on systematic construction of these codes.  These
initial results on good linear $t$-codes with $t>1$ and $q>2$ are
rather promising as they might find application in the context of
flash memories.

\bibliographystyle{IEEEtran}
\bibliography{Asy}

\end{document}